\documentclass[12pt,twoside]{article}
\usepackage{fleqn,espcrc1}

\usepackage{graphicx}
\usepackage[figuresright]{rotating}

\newcommand{\AmS}{{\protect\the\textfont2
  A\kern-.1667em\lower.5ex\hbox{M}\kern-.125emS}}

\title{Fine-tuning the basic forces of nature through the triple-alpha 
process in red giant stars\thanks{This work was partly supported by Grants  
D32513/FKFP-0242-2000/BO-00520-98 (Hungary), P13246-TPH (Austria), 
and by the John Templeton Foundation (938-COS153).}}

\author{Attila Cs\'ot\'o\address{Department of Atomic Physics, E\"otv\"os 
University, \\ P\'azm\'any P\'eter s\'et\'any 1/A, H-1117 Budapest, Hungary}, 
Heinz Oberhummer\address{Institute of Nuclear Physics, Vienna University
of Technology,\\ Wiedner Hauptstra{\ss}e 8-10, A-1040 Vienna, Austria}, 
and Helmut Schlattl\address{Max-Planck-Institut f\"ur Astrophysik, \\
Karl-Schwarzschild-Stra{\ss}e 1, D-85741 Garching, Germany}} 

\begin{document}

\maketitle

\begin{abstract}
We show that the synthesis of carbon and oxygen through the triple-alpha
process in red giant stars is extremely sensitive to the fine details of the
nucleon-nucleon (N-N) interaction. A $\pm0.5\%$ change in the strength of the
N-N force would reduce either the carbon or oxygen abundance by as much as
a factor of 30--1000. This result may be used to constrain some fundamental 
parameters of the Standard Model.
\end{abstract}

\section{INTRODUCTION}

Almost all the carbon in the Universe is produced through the triple-alpha
process \cite{Rolfs}. Although the $^8$Be nucleus, formed in the collision of
two alpha particles ($^4$He), is unbound, it lives long enough to allow the
possibility of capturing a third alpha particle to form stable $^{12}$C.
However, in order to produce enough carbon, this second reaction must be
resonant \cite{Hoyle}. The $0^+_2$ state of $^{12}$C, lying at 380 keV,
relative to the $3\alpha$ threshold, therefore plays a key role in the
synthesis of carbon. As both steps of the triple-alpha process are governed by
narrow resonances, the reaction rate is given as \cite{Rolfs}
\begin{equation}
\label{alphaa}
r_{3\alpha} \approx 3^{\frac{3}{2}} N_{\alpha}^3
\left(\frac{2 \pi \hbar^2}{M_{\alpha} k_{\rm B} T}\right)^3
\frac{\Gamma_{\gamma}}{\hbar} \exp \left(- \frac{E_{\rm res}}{k_{\rm B} T}
\right),
\end{equation}
where $M_{\alpha}$ and $N_{\alpha}$ are the mass and the number density of
the alpha particles, respectively, 
$E_{\rm res}$ and $\Gamma_\gamma$ are the resonance energy and radiative width 
of the $0^+_2$ state of $^{12}$C, respectively, and $T$ is the stellar plasma 
temperature.

\section{TRIPLE-ALPHA RATE AND N-N INTERACTION STRENGTH}

As the rate of the triple-alpha process is exponentially sensitive to
the resonance energy, even small changes in $E_{\rm res}$ can lead to 
big changes
in $r_{3\alpha}$. We studied how much $E_{\rm res}$ can change if the strength
of the N-N force is varied by a small amount. For this purpose we
used a cluster-model description of $^{12}$C \cite{our}. We performed
calculations using four different N-N forces, MHN, MN, V1, and V2
\cite{force}. Each force was tuned to give the 
experimental value for $E_{\rm res}$. Then the strengths of the forces 
were multiplied by a factor $p$,
which varied from 0.996 to 1.004. Thus $p=1.0$ gives back the experimental
resonance energy. The $E_{\rm res}$ energies as functions of $p$ are shown in
Fig.\ 1.  
\begin{figure}[!t]
\label{fig1}
\begin{minipage}[t]{8.5cm}
\includegraphics[width=8cm]{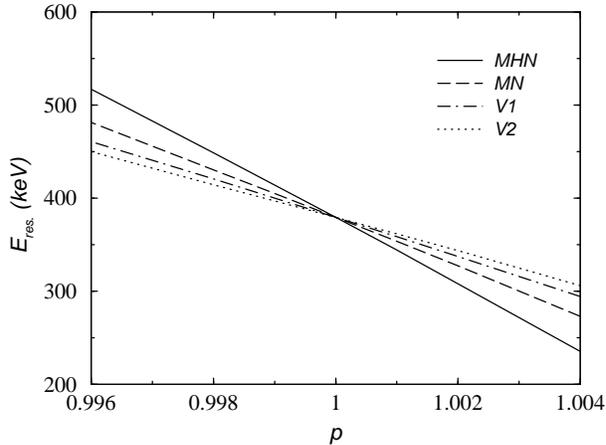}
\end{minipage}
\hspace{\fill}
\begin{minipage}[b]{7.0cm}
\caption{Resonance energy of the $0^+_2$ state of $^{12}$C as a function of
the strength parameter of the N-N forces.}
\end{minipage}
\end{figure}
One can see that small changes (less than 1\%) in the interaction strength
lead to orders of magnitude bigger changes in the resonance energy. This
effect is caused by the fact that the $0^+_2$ state lies close to the breakup
threshold, effectively behaving as a nonlinear quantum amplifier \cite{quant}.

As one can see in Fig.\ 1, the relation between $E_{\rm res}$ and $p$ 
has a force
dependence. Its origin can be traced back to the fact that the different
interactions lead to different residual alpha-alpha forces, which is the key
quantity in the nonlinear amplification phenomenon \cite{quant}. If the 
$0^+_2$ state of
$^{12}$C is correctly reproduced, then the resonance in the $^8$Be subsystem 
is underbound by the MHN force, while it is successively more and more
overbound by MN, V1, and V2. This means that from the viewpoint of the
$^{12}$C state, the residual $\alpha-\alpha$ force is too strong for MHN, and
increasingly too weak for MN, V1, and V2 (e.g., in the case of an MHN force
which reproduces the correct $^8$Be energy, the $^{12}$C state would be
overbound). This implies that the true behavior of $E_{\rm res}$ as a 
function of
$p$ is expected to be somewhere between the predictions of the MHN and MN
forces.

Using the predictions of the MHN and MN forces, we calculated the 
$r_{3\alpha}$ rates for the modified interactions, and used these rates in a
stellar model \cite{star} in order to estimate the carbon and oxygen
abundances \cite{our}. The results are shown in Fig.\ 2.
\begin{figure}[!t]
\centering
\includegraphics[width=14cm]{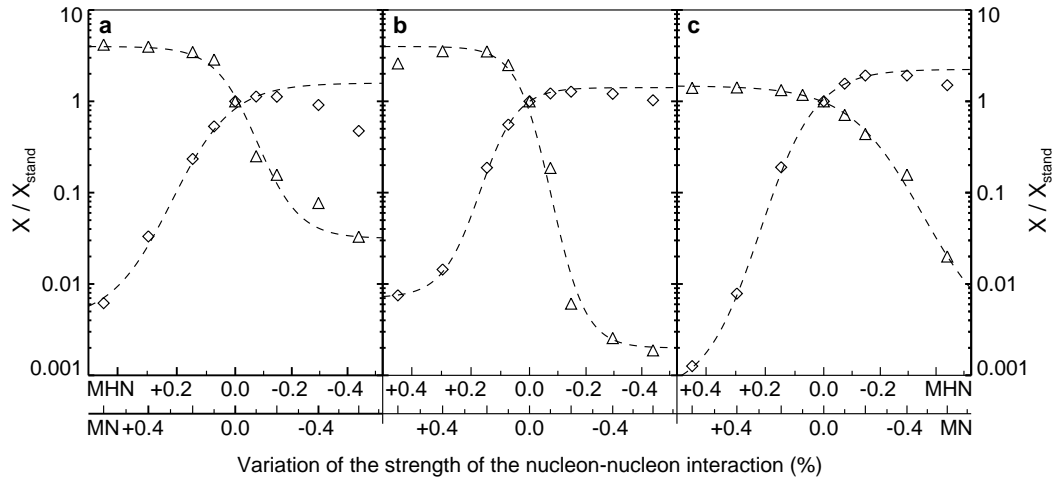}
\caption{The change of the carbon ($\triangle$) and oxygen
({\Large ${\diamond}$}) mass abundances ($X$) through variations of the strength
of the strong interaction. They are shown in panels a, b, and c for
stars with masses of 20, 5, and $1.3M_\odot$, respectively, in units
of the standard values $X_{\rm stand}$. The variations of the strength
of the strong interaction are given for the two effective N-N forces MHN
and MN. The dashed curves are drawn to guide the eye.}
\label{fig2}
\end{figure}
As one can see, the amount of carbon (oxygen) synthesized in any star is
strongly reduced if the N-N force is weaker (stronger) than the standard case.

\section{CONCLUSION}

One can see in Fig.\ 2 that a 0.5\% change in the N-N interaction
strength would lead to a Universe which does not contain an appreciable amount
of carbon or oxygen. This would make the existence of carbon-based life highly
unlikely. This very strong fine-tuning effect gives us a possibility to try to
constrain the possible values of some fundamental constants in the Standard
Model. For example, if one assumes a pion-exchange model of the strong force,
then the a smaller (larger) pion mass means a stronger (weaker) force. More
detailed analyses show that our result on the fine-tuning of carbon and
oxygen production can be used to give constraints on the possible values of 
the sum of the light quark masses and the vacuum expectation value of the 
Higgs field at the 1\% level \cite{quark}.

\end{document}